# Augmentation of VERITAS Telescopes for Stellar Intensity Interferometry

D.B Kieda*[1] for the VERITAS Collaboration,[†] S. LeBohec[1], R. Cardon[1]

[1]*Department of Physics and Astronomy University of Utah, Salt Lake City, UT 84112, USA*
*e-mail:* dave.kieda@utah.edu, lebohec@physics.utah.edu

In 2018-2019 the VERITAS VHE gamma-ray observatory was augmented with high-speed optical instrumentation and continuous data recording electronics to create a sensitive Stellar Intensity Interferometry (SII) observatory, VERITAS-SII. The primary science goal of VERITAS-SII is to perform stellar diameter measurements and image analysis in the visible wavebands on a selection of bright (m< 6), hot (O/B/A) stars. The VERITAS Collaboration has agreed to the deployment and operation of VERITAS-SII during several days each month around the full moon period when VERITAS does not perform VHE gamma-ray observations. The VERITAS-SII augmentation employs custom high-speed/low-noise focal plane instrumentation using high quantum efficiency photomultiplier tubes, and a battery-powered, fiber-optic controlled High Voltage supply. To reduce engineering time, VERITAS-SII uses commercially available high-speed (250 MS/sec), continuously streaming electronics to record the time dependence of the intensity fluctuations at each VERITAS telescope. VERITAS-SII also uses fast ( < 100 psec) data acquisition clock synchronization over inter-telescope distances (greater than 100 m) using a commercially available White Rabbit based timing solution.

VERITAS-SII is now in full operation at the VERITAS observatory, F.L.Whipple Observatory, Amado, AZ USA. This paper describes the design of the instrumentation hardware used for VERITAS-SII augmentation of the VERITAS observatory, the status of initial VERITAS-SII observations, and plans for future improvements to VERITAS-SII.

---

*Speaker

[†]http://veritas.sao.arizona.edu; for collaboration list see PoS(ICRC2019)1177.

## Introduction

The Stellar Intensity interferometry (SII) technique measures correlated light intensity fluctuations between spatially separated telescopes. The technique was first used for astronomy in the 1950s by Robert Hanbury Brown and Richard Q. Twiss, who built the twin 6.5 m diameter telescopes of the Narrabri Stellar Intensity Interferometer (NSII) [1], and used it to measure the diameters of 32 bright, nearby stars [2]. SII is relatively insensitive to both atmospheric turbulence and to telescope optical imperfections, allowing very long baselines, as well as observations at short (visible) optical wavelengths [1, 3, 4]. Astronomical observations with SII are well suited to bright, hot stars (O/B/A), and are complementary to the longer wavelength sensitivity of amplitude interferometers such as VLTI [5], CHARA [6], and NPOI [7].

During the past decade, it has become feasible to develop a superior SII observatory with imaging capabilities through the collaborative use of existing arrays of Imaging Air Cherenkov Telescopes (IACTs). IACT arrays such as VERITAS [8] are optimized for ground-based VHE gamma-ray astronomy, but also have suitable optical quality properties, sufficiently large mirror areas, and appropriate telescope spacing to create a sensitive SII Observatory. Also, telescope time is generally available during the full Moon when IACTs observatories typically do not perform VHE gamma-ray observations. This paper describes the design and installation of the instrumentation hardware used for VERITAS-SII augmentation during the 2018-2019 observing season. The article also describes the status of initial VERITAS-SII observations and plans for future improvements to VERITAS-SII. Science analysis of the inaugural year of VERITAS-SII observations is described in a separate paper at this conference [9].

## The VERITAS IACT Observatory

The VERITAS Observatory [8] is an array of 4 Imaging Atmospheric Cherenkov Telescopes (IACTs) located at an altitude of 1268 m.a.s.l at the F.L. Whipple Observatory near Amado, AZ. Each telescope features an f/1.0 12 m-diameter Davies-Cotton reflector consisting of 345 hexagonal facets, resulting in 110 m$^2$ of light collecting area. The separation between nearest neighbor telescopes is approximately 80-120 m (Figure 1). The Davies-Cotton reflector provides a 4 nanosecond spread to photons arriving at the focal plane. The combination of large telescope primary mirror area, fast optics, and 100+ meter telescope baselines provides sub-milliarcsecond resolution at optical wavelengths.

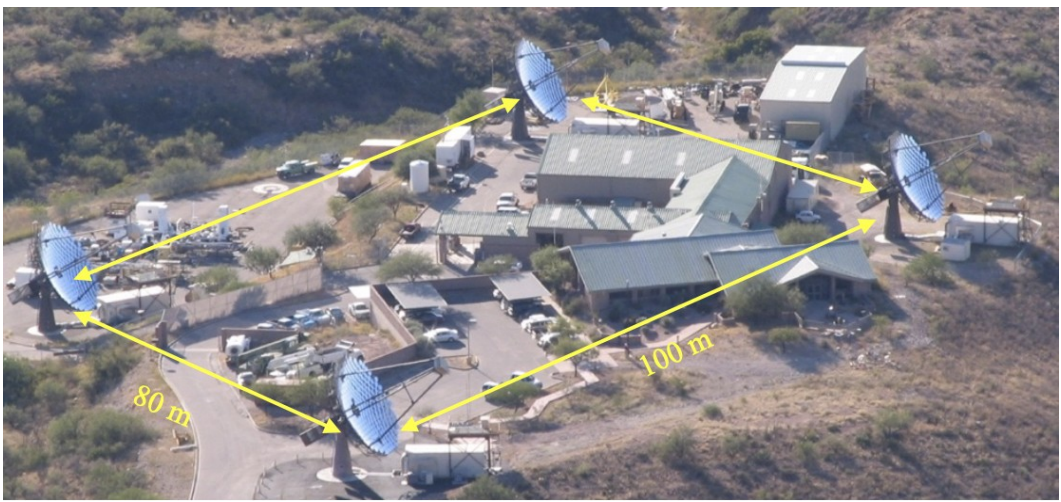

*Figure 1: Aerial picture of the VERITAS 4 telescope Imaging Air Cherenkov Telescope Observatory, including approximate telescope baseline separation distances*

**SII Focal Plane Removable Plates and Instrumentation**

The VERITAS-SII focal plane instrumentation mounts on a single removable Focal Plane Plate (FPP) that fits in front of the 499-pixel VERITAS camera (Figure 2). The FPP instrumentation includes a single Hamamatsu super-bialkali R10560 PMT [10] with a narrowband optical filter (420 nm/10 nm effective bandpass). The SII FPP instrumentation plate scale is matched to the VERITAS focal plane point spread function (PSF) as it uses the same model PMTs used for the primary VERITAS camera.

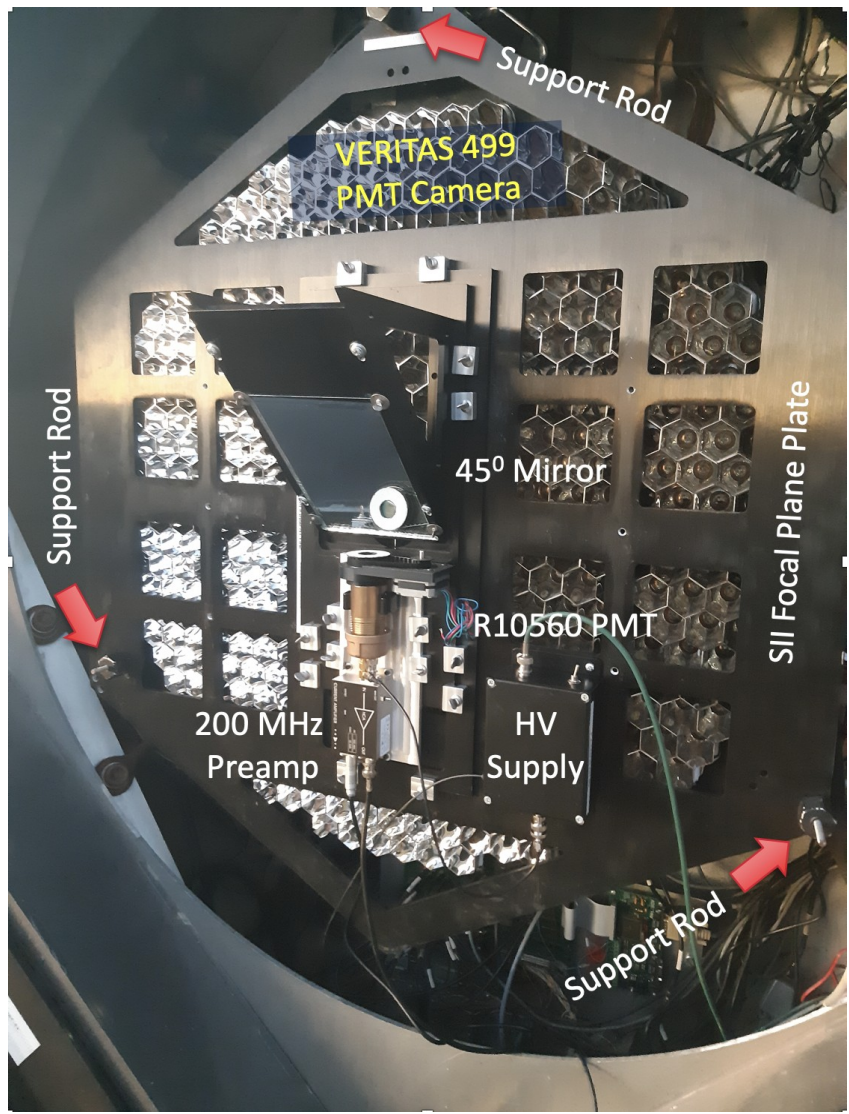

*Figure 2: Removable Focal Plane Plate (FPP) used for VERITAS-SII. The metal FPP is mounted on three threaded rods (highlighted with red arrows) in front of the VERITAS 499 pixel PMT camera. The FPP supports the 45° mirror, the R10560 PMT, the fiber optic controlled HV supply, and a 200 MHz preamplifier. The signal and control cables for the instrumentation (near the bottom) are quickly connected/disconnected when adding/removing the FPP.*

The PMT gain is set by a battery-powered, fiber-optic controlled High Voltage (HV) supply [11]. The HV setting and power is controlled through pulse width modulation of a fiber optic transmitter. The HV supply battery is permanently mounted in the VERITAS camera and is re-charged during the daytime. The output of each PMT is amplified by a high speed (200 MHz bandwidth) transimpedance preamplifier which drives a 45 m long double-shielded RG-223 co-axial cable. The coaxial cables are routed along the VERITAS telescope quadrapod arms and the optical support structure, eventually terminating in the SII data acquisition system, located in the telescope electronics trailer next to each VERITAS telescope. Special care is made to ensure the SII focal plane instrumentation is electrically isolated from any telescope grounding to minimize pickup of local RF noise.

The removable FPPs are lightweight and can be easily installed or removed in less than 10 minutes per telescope. The FPPs are designed to be securely mounted with no modifications or changes to the VERITAS cameras.

**Narrowband Optical Filter Effective Bandwidth**

The starlight from the VERITAS mirror facets reflects off a 45° front surface mirror mounted onto the FPP. At the focus, an optical diaphragm slightly larger than the expected physical PSF collects the starlight and helps minimize stray light., The light passes through a Semrock FF01-420/5-25 interferometric narrow-band filter with a center wavelength of $\lambda$ = 420 nm and a bandwidth of $\Delta\lambda$ = 5 nm at normal incidence [12]. Generally, narrow-band optical filters require collimated light. However, the Semrock filter coating exhibits a high refractive index ($n$ = 2.38), which minimizes the effect of non-normal incident photons. A simulation was used to calculate the shift in optical bandpass wavelength for the VERITAS f/1.0 mirror optics by weighting the angular distribution of incoming photons by the relevant mirror area at each angle of incidence. The resulting incidence angle distribution was then used to calculate the distribution of center bandpass wavelengths through convolution of the photon arrival distribution using the standard formula describing the shift in center bandpass wavelength with incidence angle[11]. The resulting calculation indicates an effective bandwidth of $\Delta\lambda \approx 10$ nm about a ~415 nm center wavelength, with reduced light transmission compared to normal incidence.

**VERITAS-SII Data Acquisition System**

A standalone National Instruments (NI) PXIe data acquisition system is deployed at each VERITAS telescope to enable local data recording (Figure 3). Each SII data acquisition system uses an NI PXIe crate with a high-speed PXIe backplane (3 GB/sec). The crate holds a high-speed PXIe controller, high-speed interface (PXIe-8262) connected to a high speed (700 MB/sec) RAID disk (NI-8265). Each PMT signal is continuously digitized at 250 MS/s by an NI FlexRio Module (NI-5761). The 250 MS/s digitization rate matches the 4 nsec isochronicity of the VERITAS optics. The digitized data from each PMT is scaled and merged into a single continuous data stream by a Virtex-5 FPGA module. The data stream is transported across the high-speed PXIe backplane and recorded to the RAID disk. The NI PXIe data acquisition modules are all commercially available, and standard software developed under the LabVIEW programming environment is used to control the data acquisition.

**FPGA Clock Timing Synchronization**

The intensity signals recorded between separated telescopes must be synchronized to within a small fraction of the sampling period. If this is not accomplished, the photon coherence signal will be smeared across multiple time samples, effectively degrading the electronic bandwidth and thus the achievable SNR. The distributed data acquisition system incorporates synchronization of each telescope's DAQ FPGA clock using commercially available Seven Solutions White Rabbit modules [13], synchronized through the VERITAS single-mode optical fiber network. The White Rabbit timing modules use single-mode fibers to transmit a phase-locked 10 MHz clock with a precision of < 200 psec RMS to each SII DAQ system telescope.

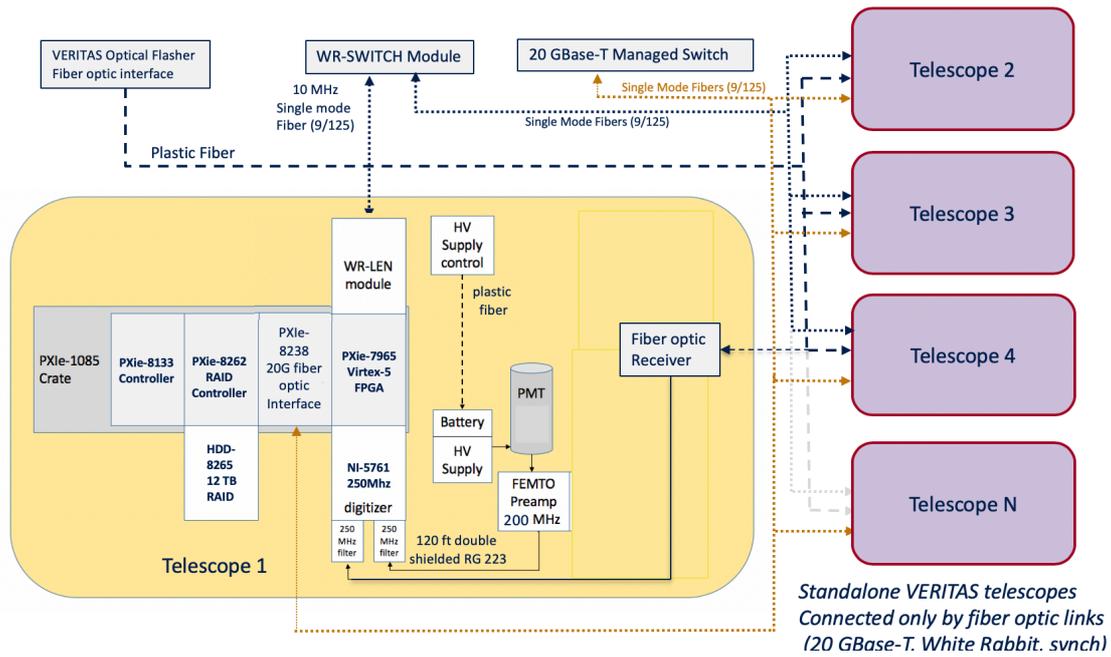

*Figure 3: Schematic drawing of VERITAS-SII DAQ architecture. Each Telescope (e.g., Telescope 1) contains a National Instruments PXIe crate hosting a system controller, RAID disk interface, PXIe-8238 10GBase-T fiber optic Ethernet controller, and FPGA/Digitizer. The Focal plane instrumentation consists of a Hamamatsu R10560 photomultiplier tube (PMT), a 200 MHz preamplifier, and a battery-powered, fiber optic controlled HV system. Each VERITAS-SII telescope is interconnected only through commercial fiber optic transceivers*

The quality of the Seven Solutions White Rabbit-based synchronization of independent telescope backplanes was experimentally verified in the laboratory. A central master clock (WR-Switch) distributed a 10 MHz reference clock via fiber optic cables to several receiver modules (WR-LEN). Each WR receiver then locked onto the central master clock and outputted a local reference clock that is subsequently phase-locked with a 125 MHz internal FPGA oscillator to generate the 250 MS/s FADC sampling rate.

A common periodic signal with 4 nsec wide pulses was then fanned out to a separate data recording channels on two physically separated SII DAQ systems. The waveform was digitized by each DAQ system and recorded on separate RAID disks. To measure the quality of synchronization between the two DAQ systems, an initial data set was recorded in with independently `free-running' clocks (i.e., no White Rabbit synchronization present). Afterward, a second data set was recorded with the two DAQ FPGA clocks synchronized by the White Rabbit system. For each data set, the cross-correlation between the two signal data streams was computed with an off-line correlator and then compared to the auto-correlation of each of the signals. As a result, the cross-correlation should peak only when the common signals are well-aligned in time. A relative drift of the DAQ FPGA sampling rates between the independent DAQ systems will generate a broad cross-correlation peak compared to the width of a single channel auto-correlation. If the sampling rates between the two DAQ systems are fully synchronized, the cross-correlation between the two independent DAQ systems will be identical to the autocorrelation of a single data channel. The resulting test indicated that complete synchronization between the two independent DAQ FPGA clocks was achieved (Figure 4).

**Two-Telescope Software Correlations**

Full cross-correlation between each SII PMT for a single night observation requires substantial computational facilities. A single 6-hour observation will generate 40 TB of data, requiring approximately 104 days of computation for a 3 GHz single-core processor using standard correlation algorithms. During Spring/Summer 2017, a technique was developed which uses each telescopes data acquisition system's high-speed disk and FPGA module to perform the 2-telescope cross-correlation in real-time. Tests at VERITAS have demonstrated the ability of the system to deliver continuous two-telescope cross-correlation at a rate equivalent to the data acquisition rate. Consequently, the nightly computational analysis and data storage requirements will be handled at the observatory site during the daytime using an FPGA module located in the same chassis as the data acquisition hardware.

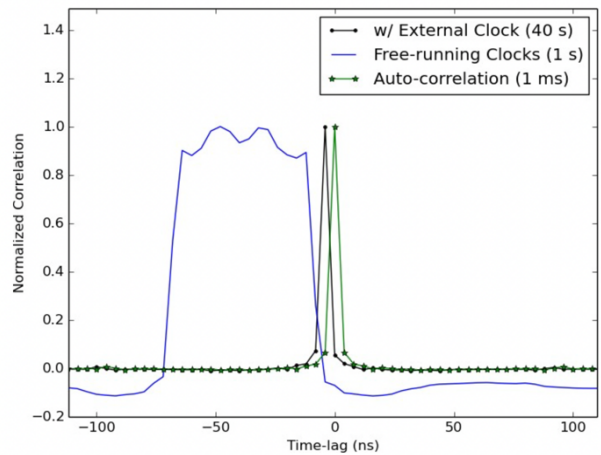

*Figure 4: Results of the White Rabbit synchronization tests performed in the laboratory. The black dotted line shows the normalized cross-correlation of a signal that was recorded to disk by independent, synchronized SII DAQ systems. Each system's 250 MHz FPGA clock is synchronized to a common 10 MHz reference clock, as described in the text. The synchronized cross-correlation is in agreement with the auto-correlation, indicating that both SII DAQ sampling rates are well synchronized. A similar test performed with the sample clocks left free-running (not synchronized) clearly shows the effect of the relative drift between the digitizers when averaged over a 1 second integration time.*

**High-Speed Data Transport Between VERITAS-SII DAQ systems**

An independent SII 20 GBase-T fiber optic data transfer network is used to provide high-speed network access to all four SII RAID data disks. This 20 Gb/sec data transfer capability allows each VERITAS-SII DAQ systems to cross-correlate their data stream with every other SII telescope data stream without having to physically move the RAID disks.

In creating the high-speed data transfer network, the VERITAS-SII DAQ system is connected by two pairs of 5/125 single-mode fiber (SMF) to a NetGear M4300-12X10 GBase-T managed ethernet switch, located in the central VERITAS building. The NetGear switch provides Link aggregation protocol (LACP) to create a single 20GBase-T link to each telescope using the two physical SMF pairs. Each SMF fiber pair uses an AXM762 SFP+ 10 GBase-T fiber optic transceiver with an identical SFP+ transceiver used to receive each SMF pair at the PXIE-8238 fiber-optic interface card in each DAQ crate. Link aggregation in each DAQ crate is handled by Intel ANS link teaming hosted under Windows 10x64.

**VERITAS SII Observation Sequence**

During a typical VERITAS-SII observation, a target star can be continuously observed for multiple hours while the source elevation is greater than > 20º. A full night's observation consists of a sequence of short (typically 20 minute) data recordings. The start and stop of each observation are triggered at each telescope by an array-synchronized one pulse per second signal, allowing easy multi-telescope data merging. An array-wide slow control system is under development to automatically coordinate VERITAS pointing with the start/stop of individual runs, as well as data transfer.

As the VERITAS telescopes track the target over 4-6 hours, the projected distance between the telescope combinations continuously changes with the movement of the star position in elevation and azimuth. The measurements of the squared visibility $|\gamma(r)|^2$ samples specific tracks across the $u-v$ Fourier image plane over East-West trajectories. By breaking up the data observation into discrete 20-30 minute intervals, the cross-correlation between individual telescopes can be associated with short track segments in the Fourier plane. For a single (1-D) track across of the Fourier image plane, a stellar diameter may be extracted from a fit to the variation of the visibility with projected telescope separation. If the Fourier image plane has been adequately sampled, a fitting process can be used to fit a putative stellar image to the observed Fourier image plane.

## 2018-2019 VERITAS-SII Observations

Two-telescope commissioning of VERITAS-SII began in Fall 2018, followed by initial two-telescope science observations at the beginning of 2019. By Spring 2019, regular two-telescope and three-telescope observations of several stars were performed under a variety of weather conditions. Observation targets included (as of March 2019) γ- Ori (9.5 hrs of observation), κ-Ori (18.2 hrs of observation), β-CMi (4.7 hrs of observation), η-Ursa Major (1.7 hrs of observation) and δ-Corvi (2.7 hrs of observation).

VERITAS-SII continued with two and three telescope observations through April 2019 and began regular four telescope operations in May 2019. A separate paper at this conference describes the VERITAS-SII analysis techniques and describes the observation of changing spatial coherence observed in two-telescope SII measurements of γ- Ori in January 2019 [9].

## Future Improvements

During Summer 2019, the VERITAS-SII system will undergo several changes that will improve the ability to use the facility for ongoing SII observations, as well as potentially increase the sensitivity of the SII system. These improvements would also provide a testbed for ideas that could be incorporated into future implementations of SII on next-generation IACT arrays, such as CTA augmentation. These improvements include:

- Improved noise rejection by using optimized preamplifier gain. This can reduce spurious RF pickup noise from local sources and substantially improve DAQ livetime.
- Installation of collapsible mirror mounts for each removable SII FPP. This improvement will remove the necessity of installing and removing the plate every day to avoid a mechanical (clearance) conflict with the camera shutter.
- Development and test of a multi-telescope slow control system which will automatically synchronize telescope target selection, pointing, data acquisition and data transfer for the full VERITAS-SII array.
- Development of a scripted two-telescope correlation data processing system to automatically process all two-telescope correlations.
- Development of post-correlation image fitting algorithms.
- Development and testing of N-telescope correlation algorithms (N>2).
- Addition of multi-wavelength instrumentation at each FPP, allowing for stellar measurements in different photometric bands.

The SII science toolkit will allow arbitrary array configurations to be simulated and will be linked to standard astronomical catalogs. The feasibility and timing of individual observations can be predicted by the observational toolkit. The toolkit will also calculate the number of hours required to achieve a specific resolution (e.g., the relative error in stellar limb darkening coefficient) as a function of stellar temperature, magnitude, observatory latitude/longitude, observation wavelength, source sky position, and lunar illumination

**Acknowledgments**

This research is supported by grants from the U.S. Department of Energy Office of Science, the U.S. National Science Foundation and the Smithsonian Institution, and by NSERC in Canada. This research used resources provided by the Open Science Grid, which is supported by the National Science Foundation and the U.S. Department of Energy's Office of Science, and resources of the National Energy Research Scientific Computing Center (NERSC), a U.S. Department of Energy Office of Science User Facility operated under Contract No. DE-AC02-05CH11231. The authors gratefully acknowledge support under NSF Grant #AST 1806262 for the fabrication and commissioning of the VERITAS-SII instrumentation. We acknowledge the excellent work of the technical support staff at the Fred Lawrence Whipple Observatory and at the collaborating institutions in the construction and operation of the instrument.

*The authors of this paper wish to recognize the key contributions by our colleague, Dr. Paul Nuñez, to the foundational aspects of this work. Dr. Nuñez passed away in an unfortunate mountain climbing accident in June 2019. We dedicate this presentation in his memory.*